\documentclass[a4paper,superscriptaddress,nofootinbib]{revtex4-1}
\usepackage{epsfig}
\usepackage{amsfonts}
\usepackage{amsmath}
\usepackage{amssymb}
\usepackage{amsthm}
\usepackage{array}
\usepackage{booktabs}
\usepackage{color}
\usepackage{graphicx}
\usepackage{hyperref}
\usepackage{multirow}
\usepackage{slashed}
\usepackage{subfigure}
\usepackage{theorem}
\usepackage{textcomp}

\allowdisplaybreaks[1]

\numberwithin{equation}{section}

\newcommand{\be}{\begin{equation}}
\newcommand{\ee}{\end{equation}}
\newcommand{\beq}{\begin{equation}}
\newcommand{\eeq}{\end{equation}}
\newcommand{\bea}{\begin{eqnarray}}
\newcommand{\eea}{\end{qnarray}}

\def\<{\left\langle}
\def\>{\right\rangle}

\begin{document}

\title{Real and virtual photons effects in di-lepton production at the LHC}

\author{Elena Accomando}
\email[E-mail: ]{E.Accomando@soton.ac.uk}
\affiliation{School of Physics \& Astronomy, University of Southampton,
        Highfield, Southampton SO17 1BJ, UK}
\affiliation{Particle Physics Department, Rutherford Appleton Laboratory, 
       Chilton, Didcot, Oxon OX11 0QX, UK}

\author{Juri Fiaschi}
\email[E-mail: ]{Juri.Fiaschi@soton.ac.uk}
\affiliation{School of Physics \& Astronomy, University of Southampton,
        Highfield, Southampton SO17 1BJ, UK}
\affiliation{Particle Physics Department, Rutherford Appleton Laboratory, 
       Chilton, Didcot, Oxon OX11 0QX, UK}
       
\author{Francesco Hautmann}
\email[E-mail: ]{hautmann@thphys.ox.ac.uk}
\affiliation{Particle Physics Department, Rutherford Appleton Laboratory, 
       Chilton, Didcot, Oxon OX11 0QX, UK}
\affiliation{Theoretical Physics Department, University of Oxford, Oxford OX1 3NP}
\affiliation{University of Antwerp, Elementary Particle Physics, Antwerp, Belgium}

\author{Stefano Moretti}
\email[E-mail: ]{S.Moretti@soton.ac.uk}
\affiliation{School of Physics \& Astronomy, University of Southampton,
        Highfield, Southampton SO17 1BJ, UK}
\affiliation{Particle Physics Department, Rutherford Appleton Laboratory, 
       Chilton, Didcot, Oxon OX11 0QX, UK}

\author{Claire H. Shepherd-Themistocleous}
\email[E-mail: ]{claire.shepherd@stfc.ac.uk}
\affiliation{School of Physics \& Astronomy, University of Southampton,
        Highfield, Southampton SO17 1BJ, UK}
\affiliation{Particle Physics Department, Rutherford Appleton Laboratory, 
       Chilton, Didcot, Oxon OX11 0QX, UK}

\begin{abstract}
We show the SM prediction of di-lepton production at the LHC where to the usual Drell-Yan production 
we add the contribution from Photon-Initiated processes. 
We discuss the effects of the inclusion of photon interactions in the high invariant mass region (TeV region) 
and their consequences on BSM heavy $Z^\prime$-boson searches.
\end{abstract}

\pacs{NN.NN.NN.Cc}

\maketitle

\tableofcontents

\setcounter{footnote}{0}

\section{Introduction}

Thanks to the recent energy improvement of the LHC machine we are now able to reach the invariant mass TeV scale,
in search for signs of heavy BSM physics.
On the other hand, exploiting the high integrated luminosity that will be collected over the next years,
we are increasing the precision of low invariant mass measurements that we can compare with high accuracy 
theoretical calculations, looking for small discrepancies from SM predictions.
In order to be successful, both these approaches require a theoretical effort to keep up with the experimental progresses.
QCD corrections at N(N)LO are now available for many processes. At this level of precision also QED (N)LO effects become 
relevant, and they have to be accounted for consistently~\cite{Martin:2014nqa}.
For some processes like di-lepton production a complete description of the QED dynamics is available~\cite{Dittmaier:2009cr}.

In this work we will focus on the effects of the inclusion of tree-level QED diagrams in SM predictions for the di-lepton final state
at the LHC. In LHC collisions protons also interact electromagnetically.
We will evaluate the contribution of photon interactions to the di-lepton final state, in comparison with the dominant 
Drell-Yan (DY) quark-initiated production.
The effect of ``quasi-real" photons ($Q^2 \simeq 0$) can be accounted for through the Equivalent Photon Approximation~\cite{Budnev:1974de},
a well known and successful procedure that exploit the dipole approximation to model the proton electromagnetic field.
More recently, PDF collaborations have released QED PDF sets, which include a component of photon within the nucleons.
Effectively in this description the photon is treated as any other parton, and we can use those QED PDFs to explore 
the effects of ``real" photons ($Q^2 = 0$), as they now resolved.

The importance of the inclusion of Photon-Initiated (PI) terms have been discussed in the literature concerning different 
final states~\cite{Accomando:2016tah, Harland-Lang:2016kog, CarloniCalame:2007cd, Alioli:2016fum, Bourilkov:2016qum, Bourilkov:2016oet, Dyndal:2015hrp, Pagani:2016caq}, 
and their effects are now accounted for in the experimental analysis~\cite{CMS:2016abv, Aaboud:2016cth}.

\section{Real and virtual photons}
\label{sec:photons}

We are interested in evaluating the effects of diagrams with two photons in the initial state, and two charged leptons in the final state.
The process $\gamma\gamma \rightarrow \ell^+\ell^-$ receive contributions from the $t$- and $u$-channel exchange of 
a charged lepton.

As mentioned, QED PDF sets can be invoked to evaluate the contribution of those diagrams.
Some of these PDF sets, such as NNPDF3.0QED~\cite{Ball:2014uwa}, xFitter\_epHMDY~\cite{Giuli:2017oii}, 
CT14QED\_inc~\cite{Schmidt:2015zda} and LUXqed~\cite{Manohar:2016nzj}, are inclusive, meaning that the elastic component resulting 
from interactions involving virtual photons is included, while some other sets, such as CT14QED and 
MRST2004QED~\cite{Martin:2004dh} are inelastic, that is the elastic component is subtracted off.
The procedures and some of the assumptions adopted by the various PDF collaborations in the fitting of Deep-Inelastic Scattering (DIS) 
and possibly LHC data, can be quite different. 
For an exhaustive description of each PDF collaboration procedure we refer to the associated literature. 

Inelastic sets can be used to separately evaluate the three terms results from two real photons, one real and one 
virtual photons, and two virtual photons interactions, denoted respectively as Double-Dissociative (DD), 
Single-Dissociative (SD), and pure EPA terms. 
We calculate these terms using~\cite{Accomando:2016ehi}:

\begin{eqnarray}
 \label{eq:integration}
 \frac{d\sigma_{DD}}{dM_{\ell\ell}} & = & \iint dx_1 dx_2 \frac{\left|\mathcal{M}(\gamma\gamma\rightarrow \ell^+\ell^-)\right|^2}{32\pi M_{\ell\ell}} f_\gamma(x_1, Q)f_\gamma(x_2, Q) \label{eq:double_diss}\\
 \frac{d\sigma_{SD}}{dM_{\ell\ell}} & = & \int dQ^2_1\iint dx_1dx_2 \frac{\left|\mathcal{M}(\gamma\gamma\rightarrow \ell^+\ell^-)\right|^2}{32\pi M_{\ell\ell}} N(x_1,Q^2_1)f_\gamma(x_2, Q)+(x_1\leftrightarrow x_2) \label{eq:single_diss}\\
 \frac{d\sigma_{EPA}}{dM_{\ell\ell}} & = & \int dQ^2_1 \int dQ^2_2 \iint dx_1dx_2 \frac{\left|\mathcal{M}(\gamma\gamma\rightarrow \ell^+\ell^-)\right|^2}{32\pi M_{\ell\ell}} N(x_1,Q^2_1)N(x_2,Q^2_2) \label{eq:EPA}
\end{eqnarray}

\begin{figure}[h]
\begin{center}
\includegraphics[width=0.45\textwidth]{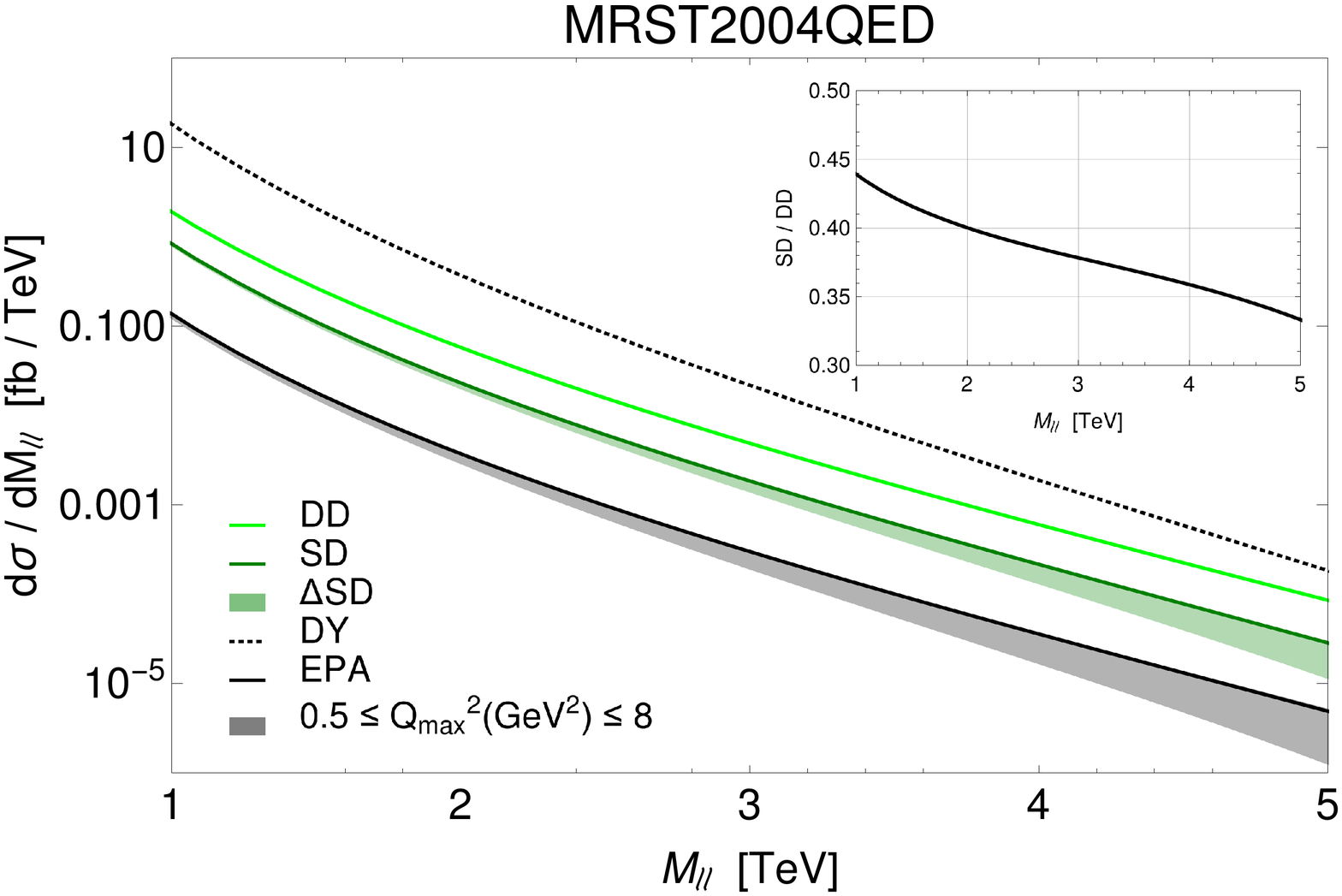}{(a)}
\includegraphics[width=0.45\textwidth]{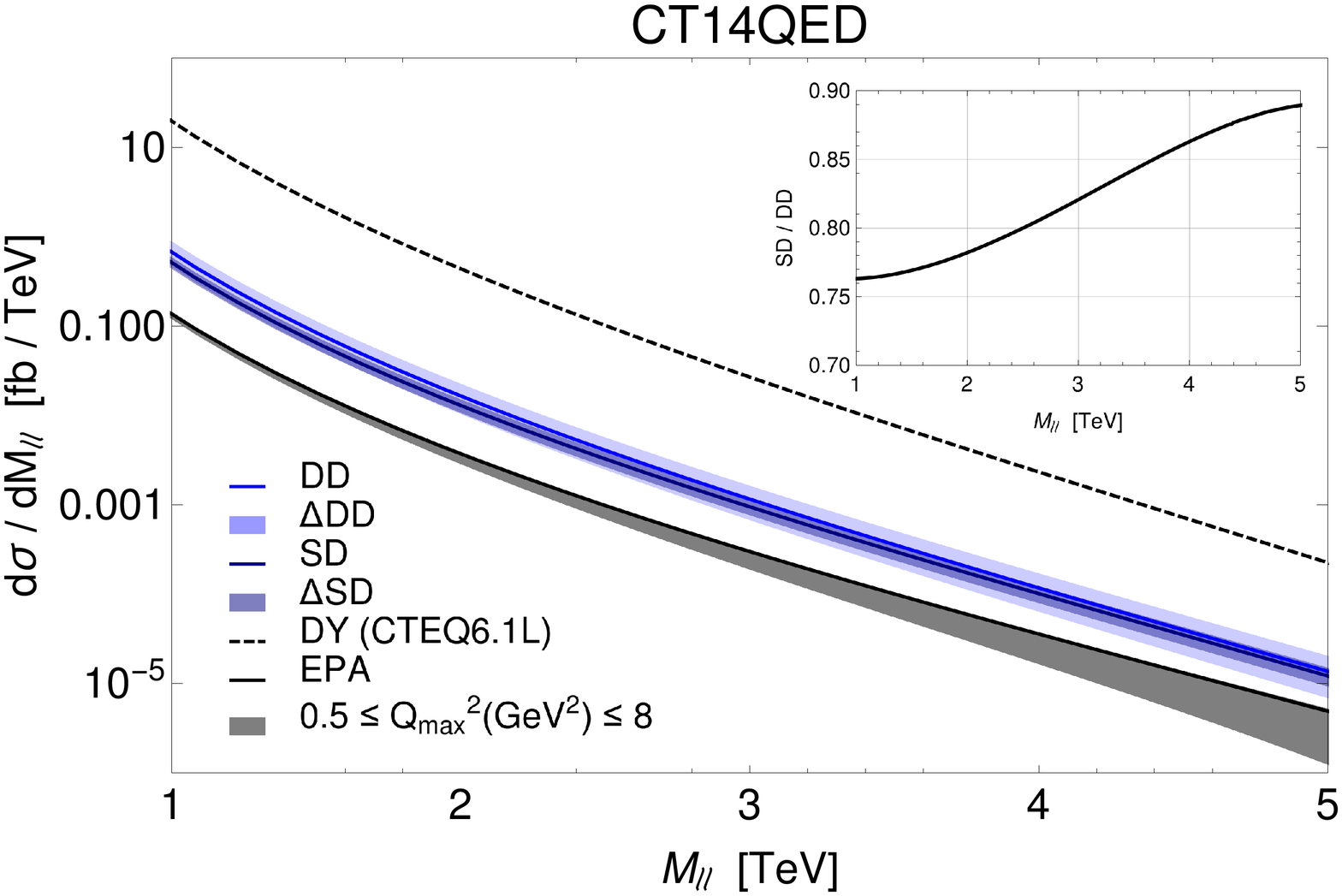}{(b)}
\caption{Individual photon-induced contributions to the di-lepton spectrum at the LHC@13TeV. 
From bottom to top, the dashed black line represent the DY prediction, the light and dark coloured 
curves represent the DD and SD results respectively, while the solid black line represents the pure EPA term.
The shaded areas include the systematic uncertainties. The inset plots show the ratio between the SD and the DD results.
We show results for the (a) MRST2004QED set and (b) the CT14QED set.}
\label{fig:PI_EPA_single}
\end{center}
\end{figure}

In eqs.~(\ref{eq:single_diss}) and~(\ref{eq:EPA}), $N(x,Q^2)$ represent the spectrum of virtual photons.
Its derivation is given in~\cite{Budnev:1974de}, while a more recent treatment, including the numerical values for the EPA 
parameters that we have used in this analysis, is available in~\cite{Piotrzkowski:2000rx}.
The integration over the virtual momentum $Q^2$ is constrained between $Q^2_{min}$ that is kinematically determined, 
and $Q^2_{max}$ that is arbitrary, and taken to be varied around the fixed value at $Q^2_{max} = 2~\rm{GeV}^2$.
Results for the three terms separately in the invariant mass range of few TeV are given in fig.~\ref{fig:PI_EPA_single} 
using the MRST2004QED and the CT14QED PDF sets respectively.

While the pure EPA contribution appears negligible with respect to the others, the DD and the SD results are of 
comparable size. In the inset plots we show their ratio. The SD size results of the order of 35 - 40\% of the DD term 
in the MRST2004QED picture, while using the CT14QED set we find that the SD size varies between 75\% and 90\% of the 
DD result in the examined invariant mass region.

In the plots we are also showing the systematic uncertainties, represented by the shaded areas around each curve.
The systematics coming from the virtual photon spectrum have been evaluated fixing different values for the $Q^2_{max}$ 
in the integrations of eqs.~(\ref{eq:single_diss}) and~(\ref{eq:EPA}), respectively $Q^2_{max} = 0.5~\rm{GeV}^2$ and 
$Q^2_{max} = 8~\rm{GeV}^2$ for the lower and upper error band.
The systematic uncertainties for the real photon spectrum have been calculated following the specific PDF collaboration 
adopted procedure. In particular for MRST2004QED we do not have any prescription to evaluate the uncertainties, while 
the CT14QED set is accompanied by a table of 31 PDFs, each one imposing a progressive constrain on the fraction of 
total momentum carried by the photon. From their analysis the upper bound is set to be $p_\gamma \leq 0.11\%$ 
at 68\% C.L.~\cite{Schmidt:2015zda} and following this result we have extracted central value 
and error band for the real photons contributions.

\section{Inclusive results and PDF uncertainties}
\label{sec:inclusive_and_systematics}
\begin{figure}[h]
\begin{center}
\includegraphics[width=0.45\textwidth]{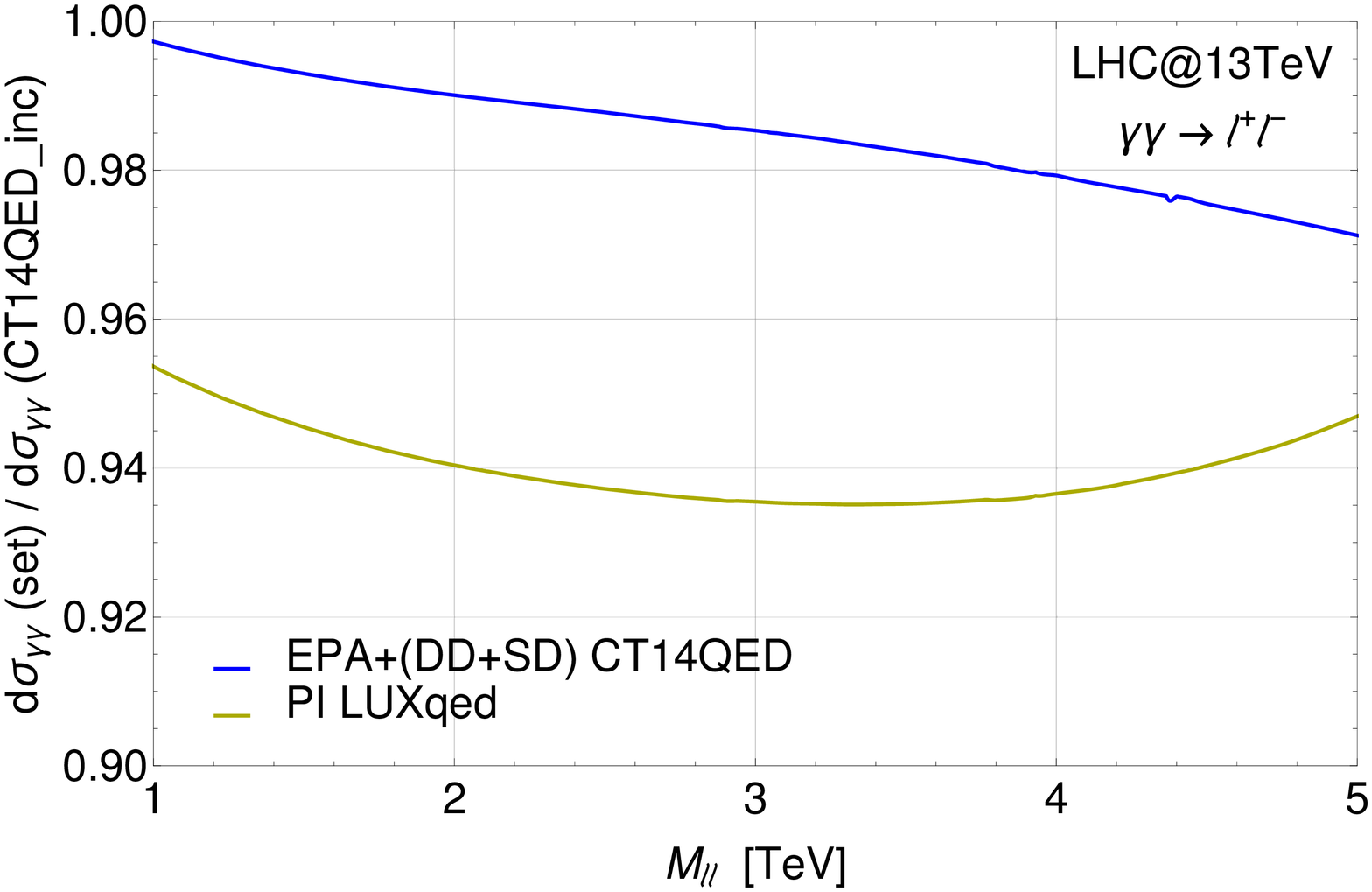}{(a)}
\includegraphics[width=0.45\textwidth]{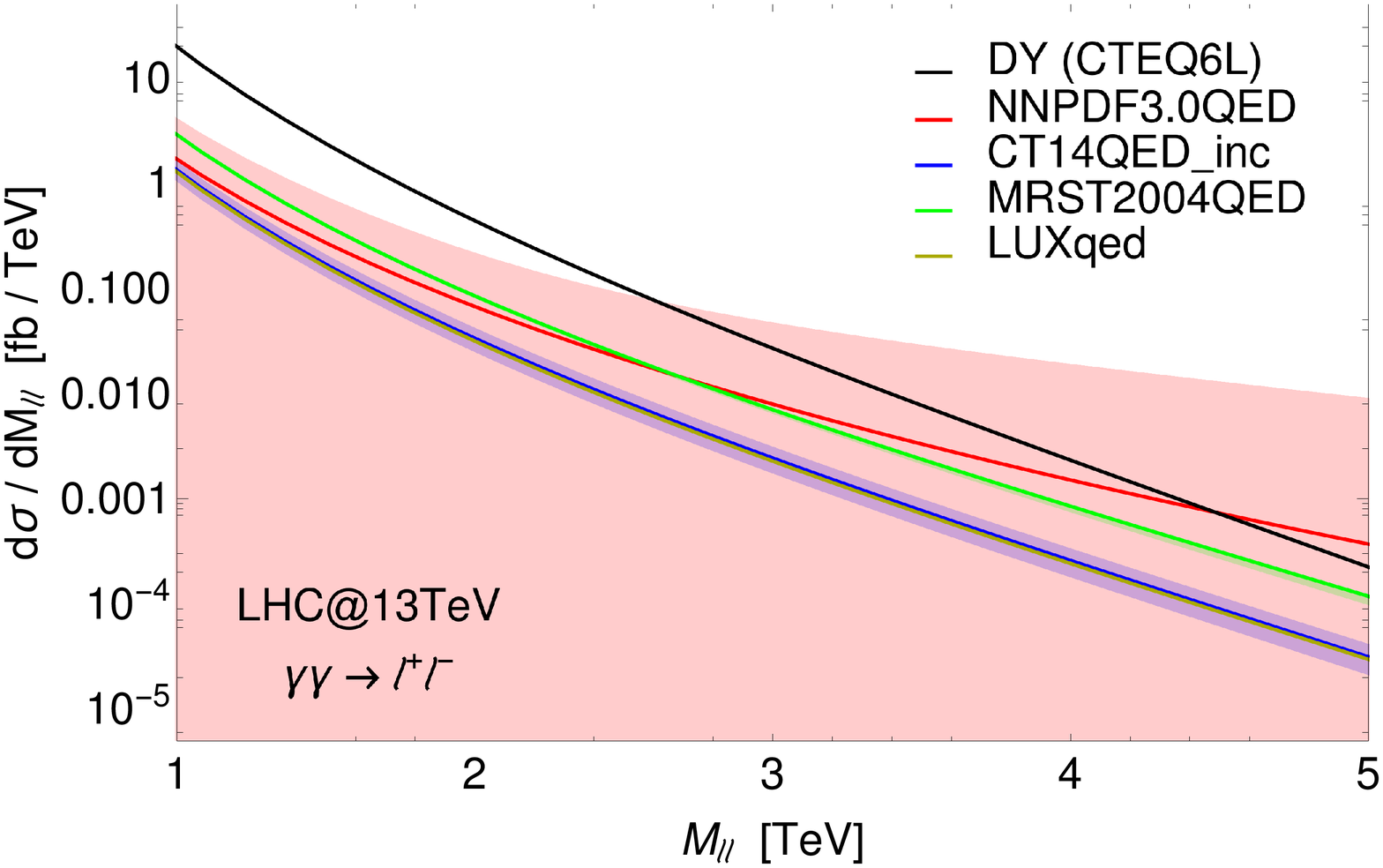}{(b)}
\caption{(a) Comparison between the full PI prediction from the CT14QED\_inc set and the LUXqed, and between the CT14QED\_inc result and the sum of the EPA, DD and SD terms obtained with the CT14QED set.
(b) Full PI prediction for different PDF sets (both inelastic and inclusive) in comparison with the pure DY term.}
\label{fig:PI_complete}
\end{center}
\end{figure}

The sum of the EPA, SD and DD terms can be directly compared with the result obtained by evaluating the DD integration 
of eq.~\ref{eq:double_diss} using inclusive sets, since they already combine both the elastic and inelastic components.
In order to verify that the separation of the various terms has been done correctly, we have compared the sum of the DD, SD and EPA 
results obtained with the CT14QED set with the inclusive result from the CT14QED\_inc set.
The result is visible in fig.~\ref{fig:PI_complete}(a) where we have plotted the ratio of those two results (blue line). 
The two results are in good agreement, the differences being $\leq 3\%$. This ensures that double counting 
effects are well under control.
In the same plot we show the comparison between two predictions for the inclusive PI results obtained with the LUXqed and the 
CT14QED\_inc sets. The two central values are in good agreement, as their difference is always $\leq 7\%$.
In fig.~\ref{fig:PI_complete}(b) we give the inclusive PI results for the various PDF sets (coloured lines) 
in comparison with the DY expectations (black line). 
The shaded areas represent the systematic uncertainties on the PI predictions.

The first thing to notice is the large difference in the error band predictions. The latter have been estimated following the 
PDF collaboration prescriptions, as already discussed for the CT14QED case.
The NNPDF3.0QED prescription follows the ``replicas" method, in this case 100 replicas contain the information on the 
systematic uncertainties. 
The LUXqed set has been released in the PDF4LHC delivery~\cite{Butterworth:2015oua}, along with a set of 100 symmetric Hessian 
eigenvectors to calculate the error.

\begin{figure}[h]
\begin{center}
\includegraphics[width=0.45\textwidth]{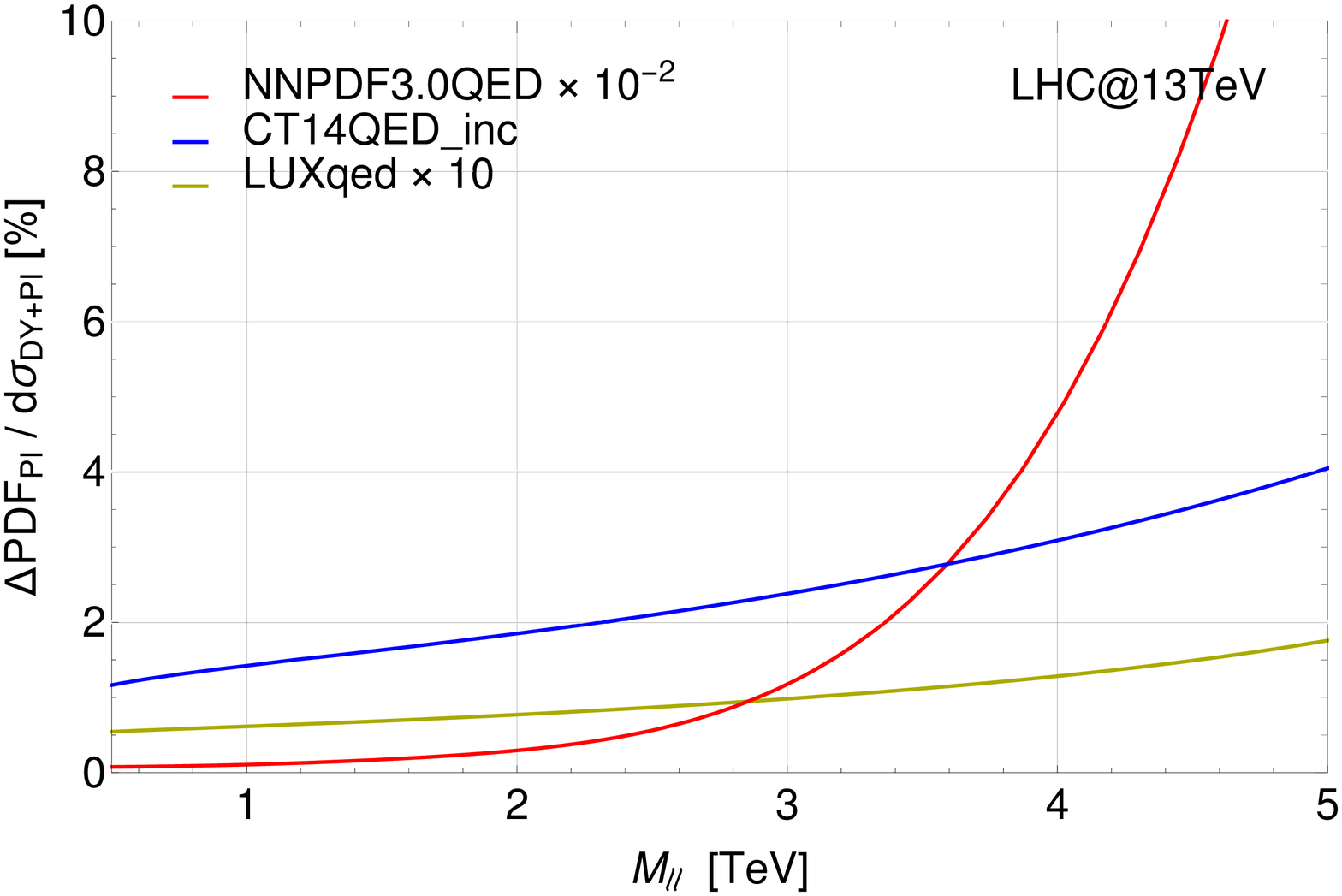}{(a)}
\includegraphics[width=0.45\textwidth]{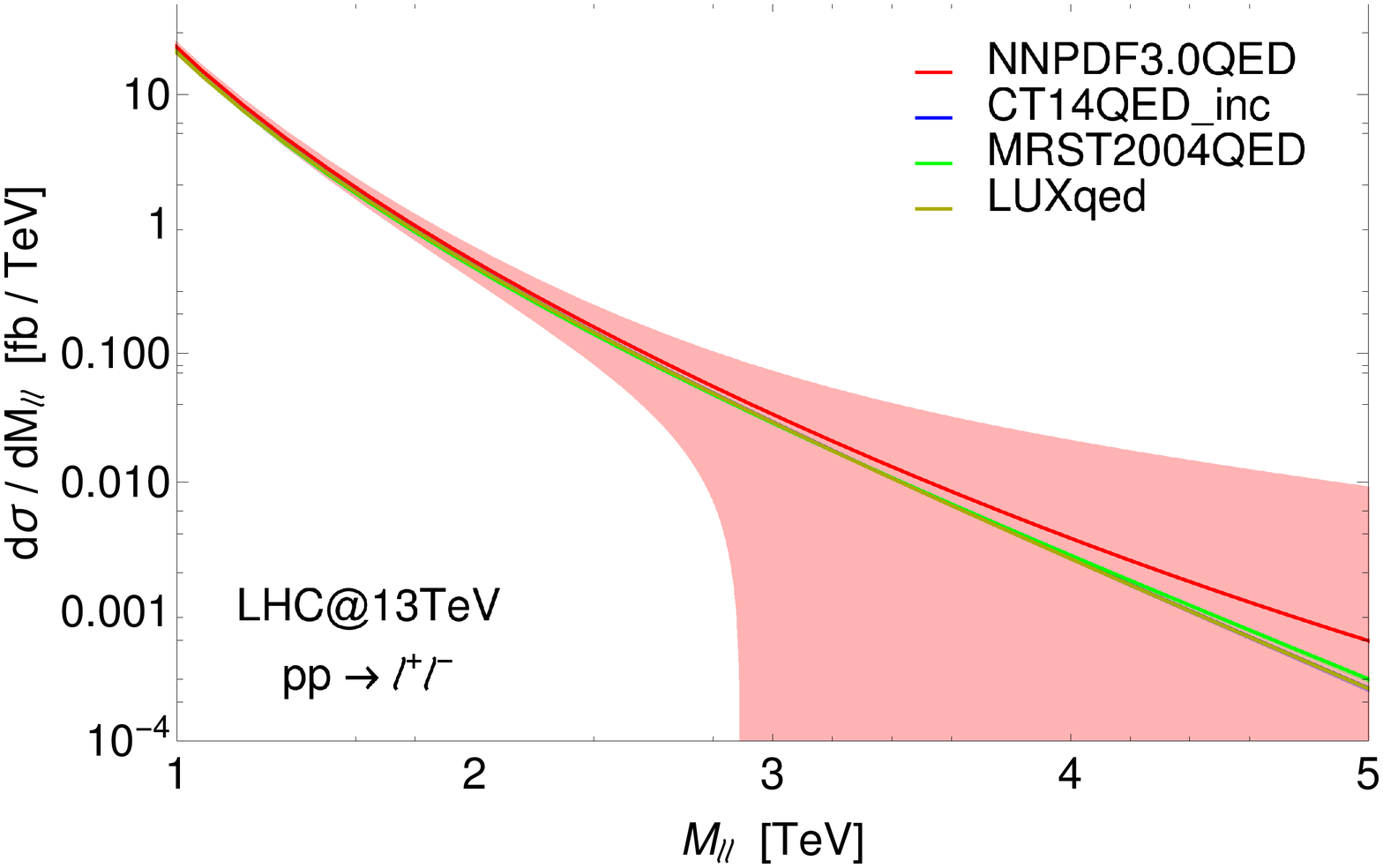}{(b)}
\caption{(a) Relative size of the photon PDF uncertainties over the complete di-lepton result. 
Note the scale factor associated to some of the curves associated to the results obtained different QED PDF sets.
(b) The di-lepton spectrum, sum of the DY and the PI central values and its PDF uncertainties as predicted by different QED PDF sets.}
\label{fig:errors_dilepton}
\end{center}
\end{figure}

The size of the errors that we have obtained varies in a wide range. 
The result is shown is fig.~\ref{fig:errors_dilepton}(a) where the three curves have different scale factors, as visible in the legend.
The most optimistic results are given by the LUXqed set, where the PDF uncertainty is of the order of 1\% along all the spectrum, 
and the central value of the PI contribution is rather small with respect to the DY term.
The most conservative scenario is given by the NNPDF3.0QED set, where the size of the systematics can be one order of magnitude larger 
than the predicted central value, which is rather large for this set.
An intermediate picture is given by the CT14QED\_inc set, where the PDF errors are between 20\% and 30\% along the spectrum.

The sum of the DY and PI central values and PDF uncertainties is given in fig.~\ref{fig:errors_dilepton}(b), where some discrepancies 
on the shape of the distribution are visible at high invariant masses.

\section{PI effects on BSM searches in the di-lepton channel}
\label{sec:BSM}

The results we have shown demand some caution when dealing with the interpretation of experimental analysis.
From here on we will consider the most conservative scenario given by the NNPDF3.0QED set.
The large uncertainties (and high central value) that we obtain adopting this PDF set, have considerable effect on the theoretical predictions 
that are to be compared with the experimental data.

\begin{figure}[h]
\begin{center}
\includegraphics[width=0.45\textwidth]{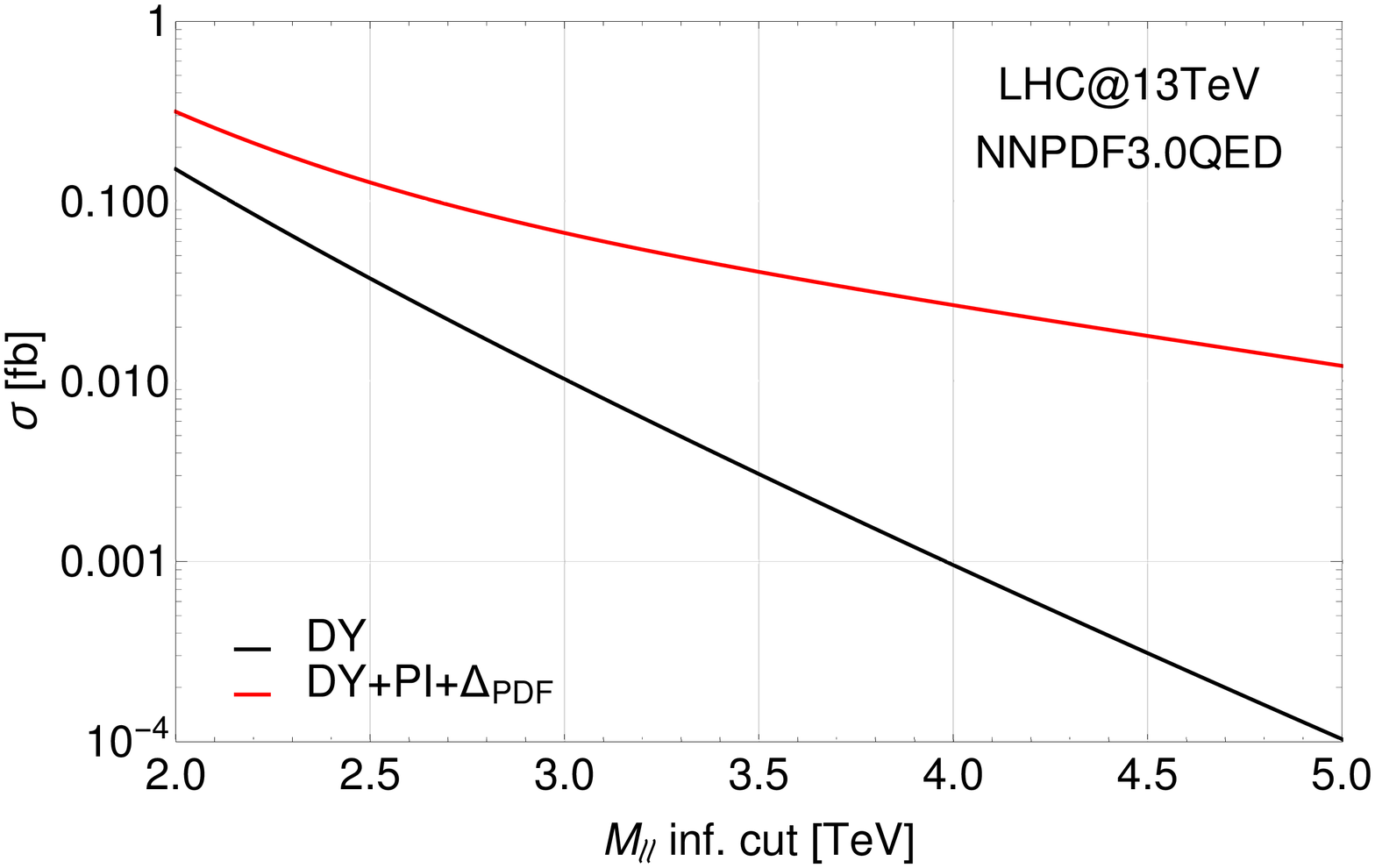}{(a)}
\includegraphics[width=0.45\textwidth]{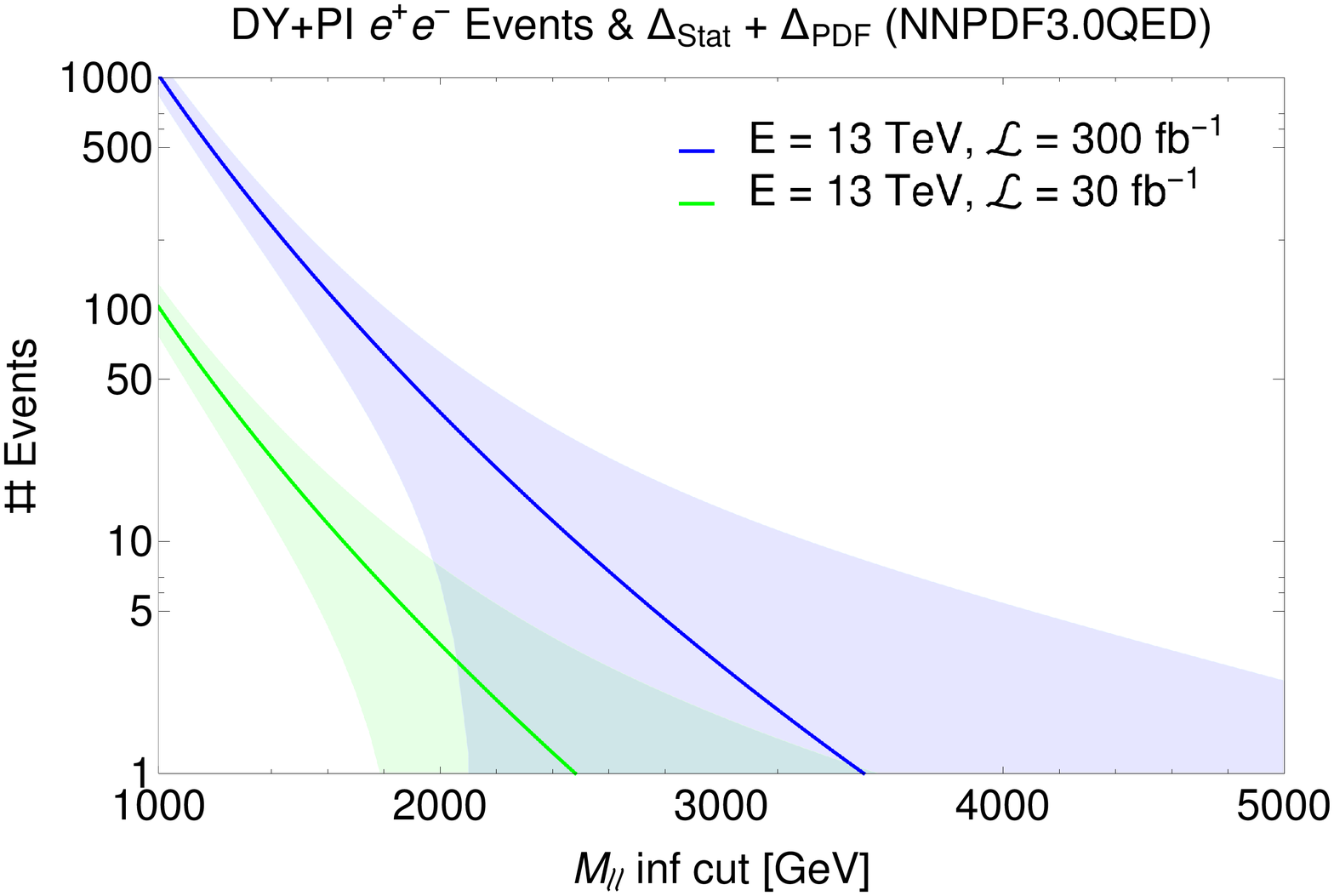}{(b)}
\caption{(a) Integrated cross section as function of the low invariant mass cut in the integration for the di-lepton channel, with and without 
the inclusion of the PI central value and PDF uncertainties.
(b) Expected number of events and its statistic and systematic error from PDF at different luminosities, 
as function of the low invariant mass cut in the integration. 
NNLO QCD corrections to Drell-Yan production~\cite{Hamberg:1990np, Harlander:2002wh} have been included according to the 
method detailed in~\cite{Accomando:2016tah, Accomando:2016ehi, Accomando:2015cfa, Accomando:2016mvz},
as well as the declared experimental acceptance and efficiency factors for the electron channel~\cite{Khachatryan:2014fba}.}
\label{fig:integrated_events}
\end{center}
\end{figure}

Especially in the high invariant mass region, the systematics show a large uncertainty on the cross section predictions.
The integrated cross section in particular receive a sizeable contribution from the PI processes inclusion.
If we consider the upper limit of the 1$\sigma$ error band, the integrated result differs up to two orders of magnitude with respect to the 
pure DY prediction. This is shown in fig.~\ref{fig:integrated_events}(a), with the two results as in the legend.

Considering the high integrated luminosity that will be collected over the years, we will soon expect to register some high invariant mass events.
For their correct interpretation, the inclusion of PI processes and uncertainties in the analysis is crucial.
As visible in fig.~\ref{fig:integrated_events}(b), because of the uncertainty on the spectrum at high invariant masses, we can observe 
events even above 5~TeV invariant mass.

We want to explore more in detail the effects of these uncertainties in BSM searches for heavy resonances in the di-lepton channel.
Neutral heavy resonances naturally appear in a variety of BSM constructions and the two leptons final state is the 
golden channel for the detection of their decays in collider experiments.
Here we have chosen two popular benchmarks~\cite{Accomando:2010fz} to study the consequences of our previous results on BSM $Z^\prime$-boson searches.
Current limits on the masses of these objects are set at few TeV~\cite{CMS:2016abv, Aaboud:2016cth}.

\begin{figure}[h]
\begin{center}
\includegraphics[width=0.45\textwidth]{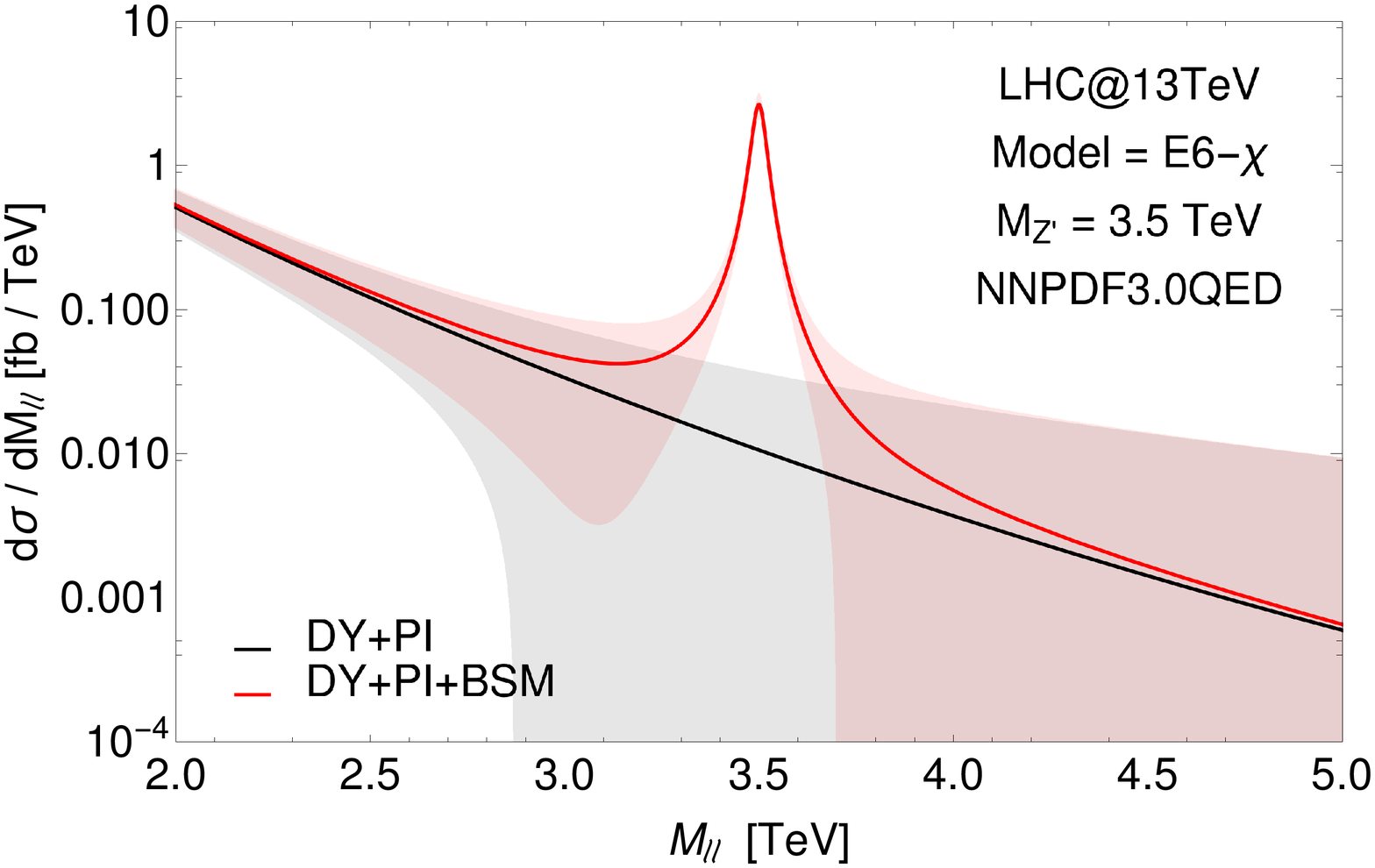}{(a)}
\includegraphics[width=0.45\textwidth]{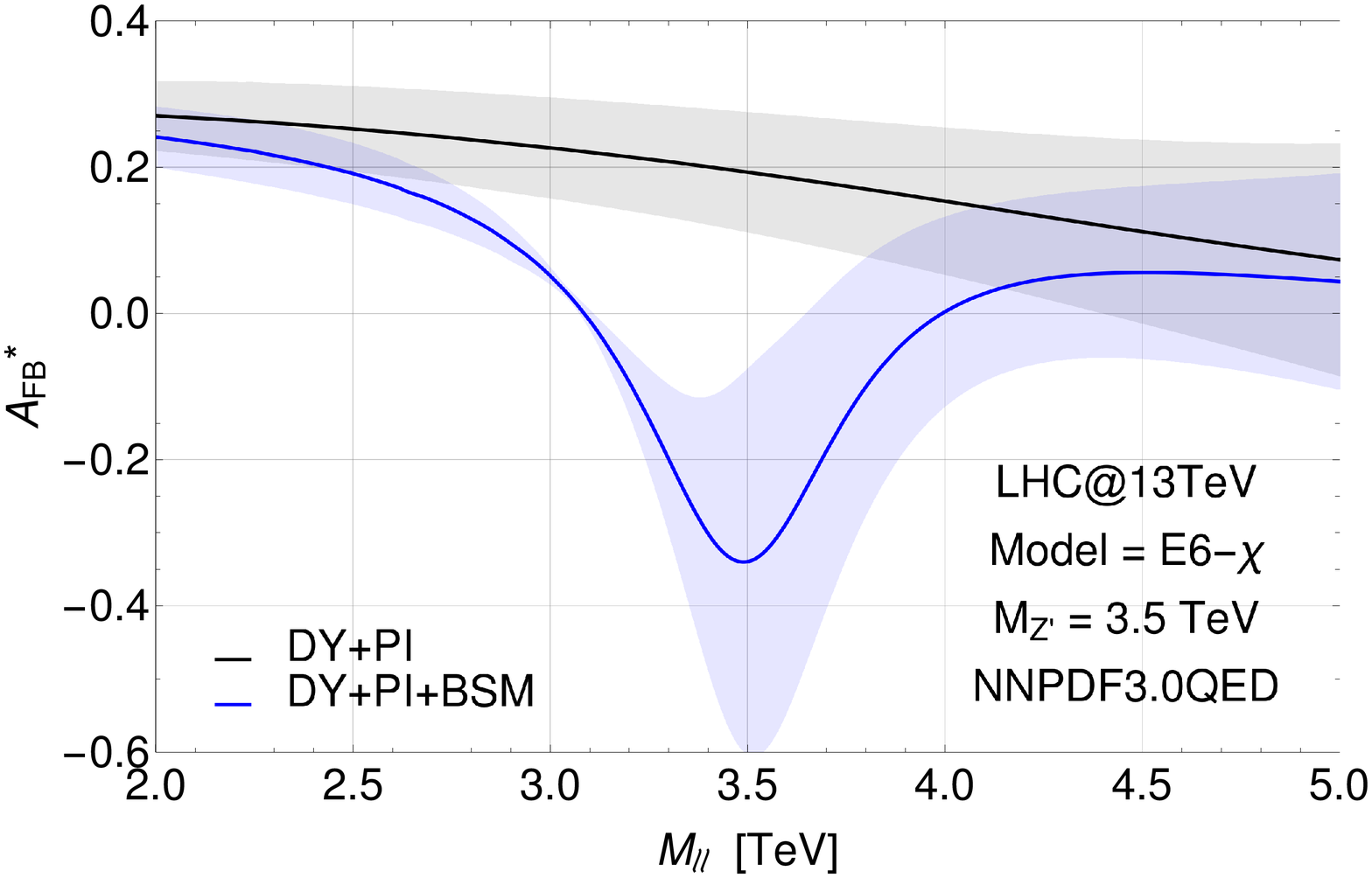}{(b)}
\caption{(a) Differential cross section and (b) $AFB^*$ distribution for a heavy $Z^\prime$ with 3.5~TeV mass as predicted by the 
$E_6^\chi$ model. The shaded areas represent the 1$\sigma$ PDF error band.}
\label{fig:E6_chi}
\end{center}
\end{figure}

To probe the narrow $Z^\prime$ case, we have considered the $E_6^\chi$. As visible in fig.~\ref{fig:E6_chi}(a) the Breit-Wigner peaked shape 
stands well above the uncertainties, thus bump searches for resonant objects do not seem to be much affected by photon PDFs errors 
in the invariant mass range under exam.
A combined analysis with the inclusion of another observable, like the Forward-Backward Asymmetry (AFB), could help in the interpretation of the results.
In the AFB observable systematics uncertainties are indeed significantly reduced~\cite{Accomando:2016ehi, Accomando:2015cfa}, thus a deviation from the flat SM predictions 
would be a clear sign of new physics. We give an example for a narrow resonance signal in fig.~\ref{fig:E6_chi}(b), again for the $E_6^\chi$ benchmark.
However, even if the large systematics from the PI result are partially cancelled, the inclusion of those terms has an effect on the SM predictions.
The PI being an angularly symmetric process, has the effects of reducing the overall AFB, thus their inclusion is essential for the correct 
extrapolations of SM background.

\begin{figure}[h]
\begin{center}
\includegraphics[width=0.45\textwidth]{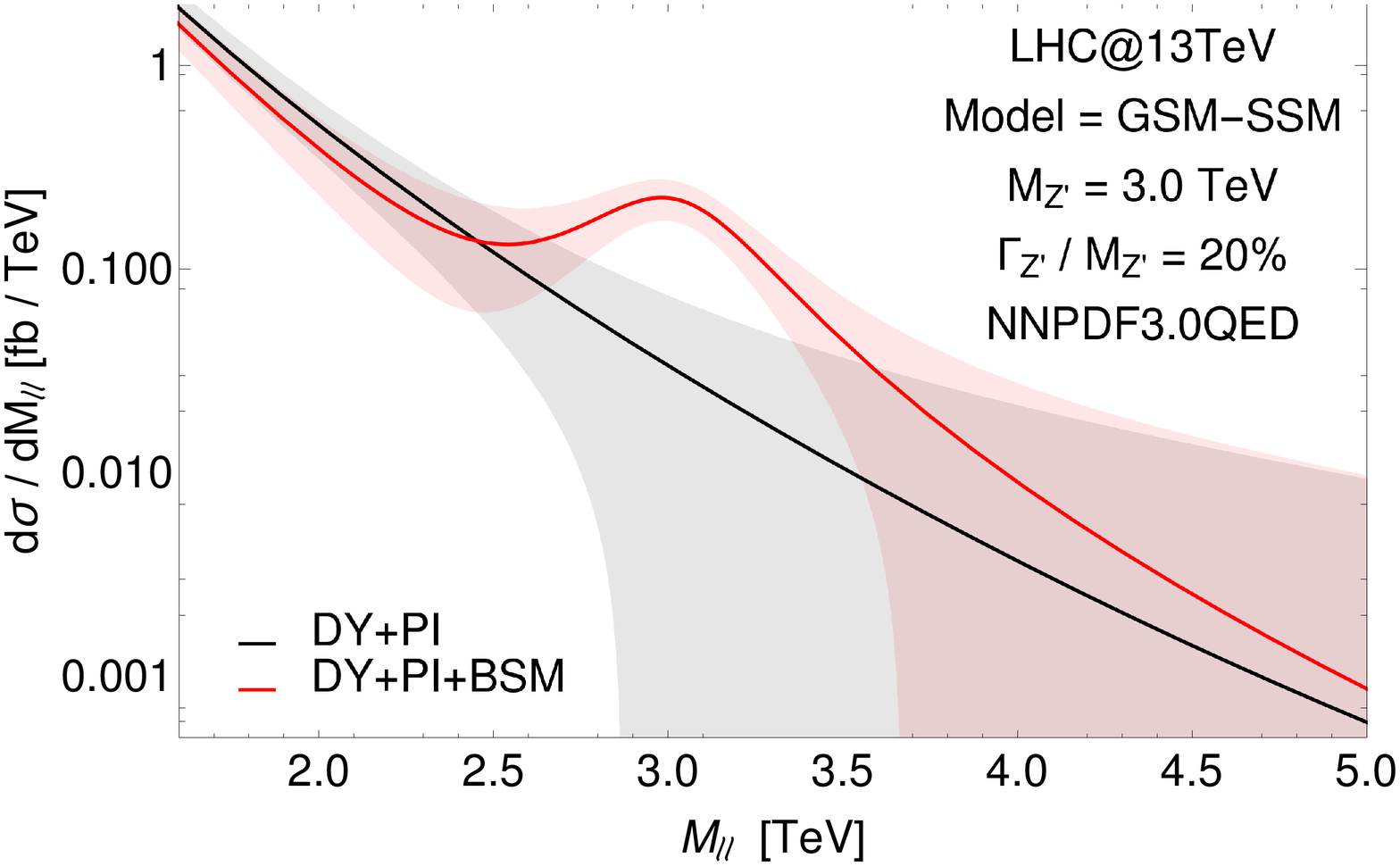}{(a)}
\includegraphics[width=0.45\textwidth]{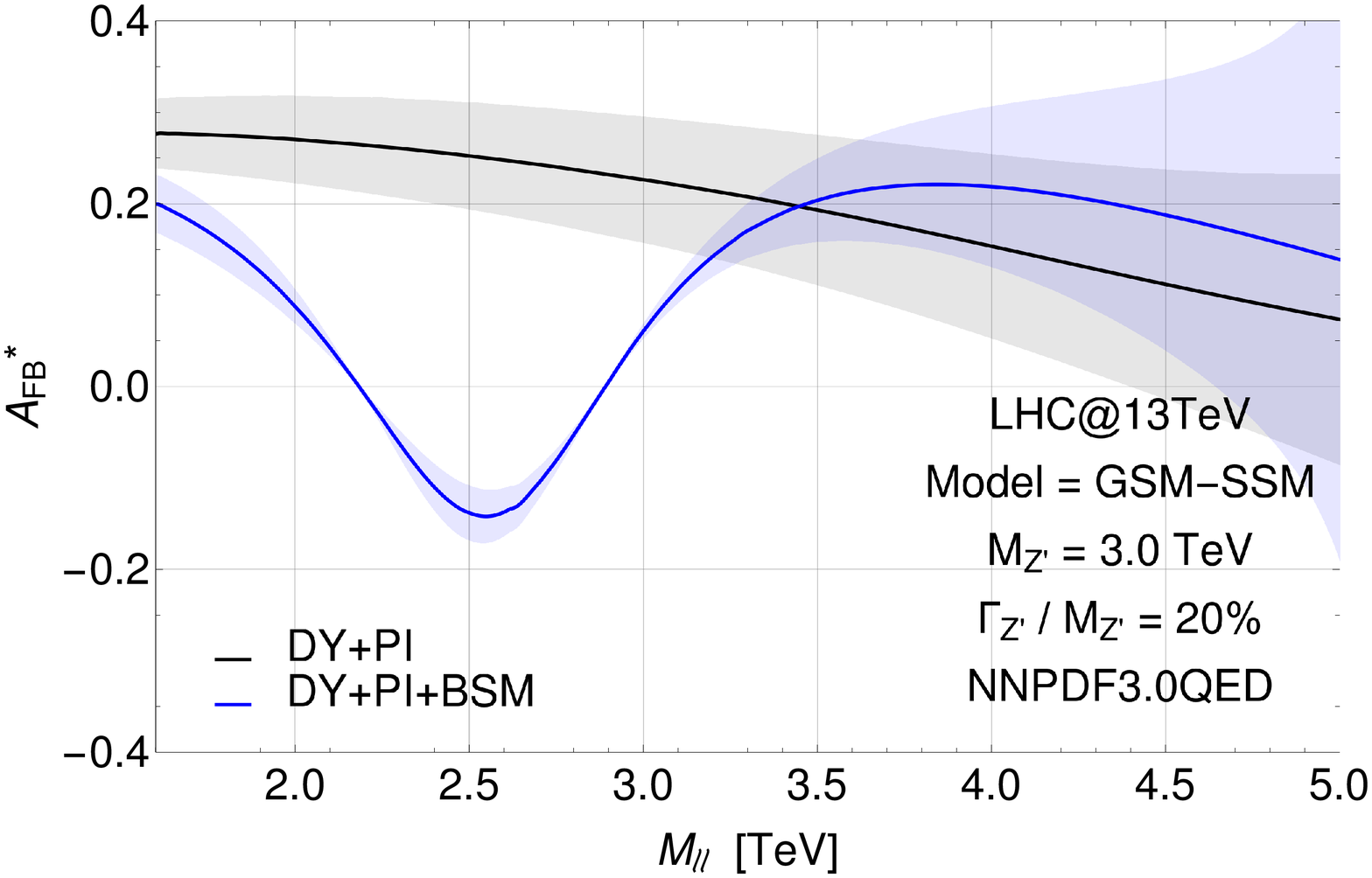}{(b)}
\caption{(a) Differential cross section and (b) $AFB^*$ distribution for a heavy $Z^\prime$ with 3~TeV mass as predicted by the 
$GSM-SSM$ model. The width of the resonance has been fixed to 20\% of its mass. The shaded areas represent the 1$\sigma$ PDF error band.}
\label{fig:SSM}
\end{center}
\end{figure}

With the purpose of exploring the PI effects on non-resonant searches, we are here considering another popular BSM model featuring a $Z^\prime$-boson,
the $GSM-SSM$ model, where the resonance width is enhanced.
In fig.~\ref{fig:SSM}(a) we show the invariant mass profile of a $Z^\prime$-boson with fixed width over mass ratio at 20\%.
This picture is quite representative of many other scenarios, like contact interactions, extra dimensions~\cite{Accomando:1999sj, Accomando:2016mvz}, 
or continuous spectra of BSM resonances, 
as in the ADD model~\cite{Mimasu:2011sa}.
As visible, now the error bands clearly reduce our sensitivity to this kind of signal. As well the typical shape of a non-resonant object 
resulting similar to a shoulder, can be confused with the tail of PI processes contribution.
Again, in this context the AFB observable can be used to corroborate the interpretations, since a typical $Z^\prime$ signal would maintain its 
shape even in the wide resonance scenario, and again the uncertainties from photon PDFs are here reduced.

\section{Conclusions}
\label{sec:conclusions}

We have discussed the contributions of PI processes to the di-lepton production channel at the LHC. 
The contributions of the new interactions have been obtained separating the effects of ``quasi-real" and ``real" photons, and they have been 
treated adopting the EPA and with the use of QED PDFs respectively.
We have computed the central values of the PI terms using different QED PDF sets and when available we have also estimated the PDF uncertainties 
along following the appropriate prescriptions.

We have discussed those results and compared with the dominant DY contribution. 
In particular at high invariant masses, deviations from the pure DY predictions can occur.
The size of these deviations, and the associated theoretical systematics, vary significantly with the different scenarios for the photon PDFs.
We have analysed the sensitivity of BSM searches for both resonant and non-resonant objects, in light of the 
previous results, adopting the most conservative QED PDF set.

Bump searches for resonant objects that follow a peaked Breit-Wigner shape are not much affected by photon interactions, while counting experiments 
for non-resonant objects would suffer a significant loss of sensitivity.
The interpretation of experimental data can be supported by introducing an extra observable as the Forward-Backward Asymmetry, particularly because 
of its favourable features concerning systematic uncertainties.

\acknowledgments
We thank A. Belyaev and A. Pukhov for discussions on the EPA. We are grateful to V. Bertone for discussions on QED PDFs and for his help with NNPDF parton distribution calculations.
This work is supported by the Science and Technology Facilities Council, grant number ST/L000296/1. 
F. H. acknowledges the support and hospitality of DESY, the University of Hamburg and 
the DFG Collaborative Research Centre SFB 676 ``Particles, Strings and the Early Universe".
All authors acknowledge partial financial support through the NExT Institute.

\bibliographystyle{apsrev4-1}
\bibliography{bib}

\end{document}